\documentclass[aps, prf, reprint, groupedaddress, nofootinbib, longbibliography, showpacs]{revtex4-1}
\usepackage{amsmath,amssymb}
\usepackage{graphics,epsfig}
\usepackage{graphicx}
\usepackage{xcolor}
\usepackage{siunitx}

\newcommand{\vct}[1]{{\mbox {\boldmath $#1$}}}

\begin{document}

\title{Structure{,} dynamics {and reconnection} of vortices in a non-local model of superfluids}
\author{Jason Reneuve}
\author{Julien Salort}
\author{Laurent Chevillard}

\affiliation{Univ Lyon, Ens de Lyon, Univ Claude Bernard, CNRS, Laboratoire de
Physique, 46 all\'ee d'Italie F-69342 Lyon, France}

\date{\today}

\begin{abstract}

{We study the reconnection of vortices in a quantum fluid with a roton minimum, by numerically solving the Gross-Pitaevskii (GP) equations. A non-local interaction potential is introduced to mimic the experimental dispersion relation of superfluid $^4\mathrm{He}$. We begin by choosing a functional shape of the interaction potential that allows to reproduce in an approximative way the so-called roton minimum observed in experiments, without leading to spurious local crystallization events. We then follow and track the phenomenon of reconnection starting from a set of two perpendicular vortices. A precise and quantitative study of various quantities characterizing the evolution of this phenomenon is proposed: this includes the evolution of statistics of several hydrodynamical quantities of interest, and the geometrical description of a observed helical wave packet that propagates along the vortex cores. Those geometrical properties are systematically compared to the predictions of the Local Induction Approximation (LIA), showing similarities and differences. The introduction of the roton minimum in the model does not change the macroscopic properties of the reconnection event but the microscopic structure of the vortices differs. Structures are generated at the roton scale and helical waves are evidenced along the vortices. However, contrary to what is expected in classical viscous or inviscid incompressible flows, the numerical simulations do not evidence the generation of structures at smaller or larger scales than the typical atomic size.}

\end{abstract}

\maketitle

\section{Introduction}

Turbulence in quantum fluids is the study of the motions induced by
a tangle of quantum vortices, which are created 
under a large-scale stirring force, or in the presence of a counter-flow generated by a
hot source
(see for instance the reviews~\cite{VinNie02,Tsu08,PaoLat11,BarLvo14,BarSkr14}). 
In practice, it can be investigated in a variety of physical systems, e.g.\ 
in cold atoms Bose-Einstein condensates~\cite{horng2008, cidrim2017}, 
superfluid $^3\mathrm{He}$~\cite{blaauwgeers2002, bradley2008}, or 
superfluid $^4\mathrm{He}$~\cite{MauTab98, zhang2005_nature}.
In this paper, we focus on the case of superfluid $^4\mathrm{He}$, which is
obtained when liquid $^4\mathrm{He}$ is cooled at temperatures 
below $T_{\lambda} \approx \SI{2.17}{K}$ (at saturated
vapor pressure).
At finite temperature,
the fluid is made up of a
mixture of two components, one being classical and viscous, governed by the
incompressible Navier-Stokes equation, and the other one being inviscid,
compressible and potential with localized and quantized singularities (i.e.\ 
quantum vortices) hosting the rotational motions. These two components interact
in a subtle way through the friction of these vortices onto the viscous
component, a phenomenon that allows the decay of fluctuations of the quantum
component without thus the action of viscosity. Macroscopically, i.e.\ at scales
larger than the dissipative scale of the classical component, statistics of
velocity fluctuations in this mixture look very similar to the ones observed in
classical three-dimensional turbulence, as depicted in the phenomenology of
Kolmogorov~\cite{Fri95}. This includes the fine scale structure of turbulence,
such as the power-law decrease of the velocity spectrum and higher-order (i.e.\ 
intermittent) properties~\cite{MauTab98,AbiBra98,SalBau10,VarGao18}, scale-energy
transfers (i.e.\ the skewness phenomenon)~\cite{SalCha12}, and also the global
behavior at large scales~\cite{SaiHer14}. Even if some differences have been
highlighted between quantum and classical turbulence~\cite{PaoFis08,WhiBar10}
at the level of an isolated quantum vortex, it is thus tempting to consider that
at a finite temperature below $T_\lambda$, the two components are locked at
each others, implying that quantum vortices self-organize, forming structures
(i.e.\ bundles) such that the overall locally averaged vorticity field of the
superfluid component resembles to the one observed in classical 
turbulence~\cite{VinNie02,Tsu08,PaoLat11,BarLvo14,BarSkr14}. At smaller scales, typically
below the mean inter-vortex distance, {a decoupling of the two components is nonetheless expected}~\cite{salort2011}, superfluid velocity
fluctuations being governed by other phenomena such as Kelvin waves propagation
along vortex cores~\cite{KozSvi04,LvoNaz07}.

Such scales are difficult to access experimentally, thus from a modeling point
of view, it is tempting to study the collective effects of a population of
localized singularities hosting a distributional repartition of vorticity, in
particular in interaction with an exterior (classical) velocity field.  A
popular method to study the interaction between these vortices {is given by the Vortex Filament (VF) model, where the velocity field induced by a vortex is described by the Biot-Savart law. In the case where the vortex core size is negligible compared to its local curvature, it is interesting to consider the Local Induction Approximation (LIA) \cite{Sch85,Sch88} of the VF model, that takes only into account the induced velocity field of a local portion of the vortex. Such an
approach can then be} generalized to take into account non local effects induced
by the whole vortex~\cite{AdaFuj10,BagBar12}, in order to study the dynamics of an
ensemble of vortices and the implication of non local effects. In this context,
a phenomenon of tremendous importance is vortex reconnection, that allows for
dissipation and a possible change in the macroscopic distribution of these
vortices, that has to be put by hand at this stage. 

An alternative approach devoted to the dynamics of vortices and their
interaction, that might be of some interest in the improvements of the
aforementioned discrete approaches, neglecting the coupling with a normal
component and assuming vanishing temperature $T=0$, is given by the description
and evolution of the order parameter of the superfluid in terms of fields, as
it is proposed by a partial differential equation known as the Gross-Pitaevskii
(GP) equation~\cite{PitStr03}. Contrary to the approach based on the LIA, the
GP equation {naturally includes} vortex reconnection, without
thus asking for further modeling steps. This approach has been studied 
{extensively} in the literature~\cite{KopLev93,NorAbi97}. In this approach,
that allows to study some global behavior of an assembly of vortices at scales
of the order of the inter-vortex distance, their very internal structure is neglected and the two
body interaction usually adopted is of localized (i.e.\ distributional) type.
The purpose of this article is the numerical study of a non-local version of the
GP equations that allows to reproduce a realistic internal structure of these
vortices and observe and quantify its implication on vortex reconnection, and
on possible {wave packet} propagation along their cores. This can be done while
introducing in the interaction potential a physical length scale $a$
representing the typical size of a $^4\mathrm{He}$ atom.

\section{A non local model of superfluids including the roton minimum, and its calibration}

\begin{figure}[htbp]
\epsfig{file=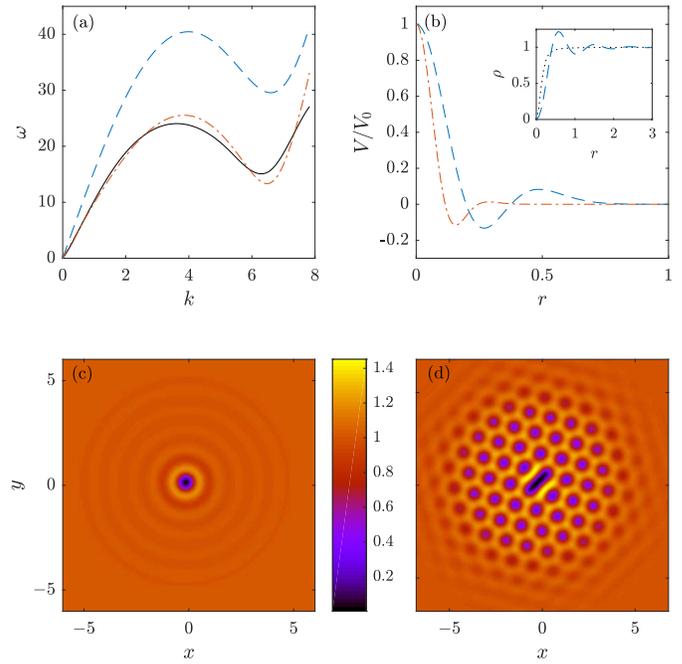,width=8.8cm}
\caption{(a): Dispersion relations of $^4\mathrm{He}$ and its model as given by the
non-local GP equation (Eq.~\ref{eq:GPNL}):  experimental results from neutron
scattering experiments (black line, see Ref.~\cite{DonBar98}), its fit as
proposed in Ref.~\cite{BerRob99} (red dashed-dotted line), {and} our current model
(Eq.~\ref{eq:InterPot}) with a raised roton minimum so as to avoid
crystallization as explained in the text (blue dashed line). (b):
Corresponding interaction potentials $V(r)/V_0$ entering in Eq.~\ref{eq:GPNL},
with $V_0=21.96 \,\mbox{eV}$ for the model of Ref.~\cite{BerRob99} $V_0=0.15
\,\mbox{eV}$ for our present model (Eq.~\ref{eq:InterPot}), used to obtain the
theoretical dispersion curves of (a). We display in the inset the radial
density profile of the stationary solution corresponding to the local GP
equation (Eq.~\ref{eq:GPL}) with a black dotted line, and the corresponding
profile in the non-local case (Eq.~\ref{eq:GPNL}) with a blue dashed line. {(c)
and (d): Color plots of the distribution of density, i.e. $|\psi|^2$, of the
stationary vortex solution obtained as the solution of the relaxation problem given in Eq.~\ref{eq:2Drelax}. (c): with our raised
roton gap (Eq.~\ref{eq:InterPot}) so as to avoid crystallization. (d): with the potential proposed in Ref.~\cite{BerRob99}, which leads to a crystallization event.} \label{fig:DispRelStatSol}}
\end{figure}

At zero temperature $T=0$, to the lowest approximation we can consider the
superfluid under study to be described by a scalar wave function $\psi$, i.e.\ an
order parameter, which is space and time dependent. Henceforth, we consider
dimensionless coordinates $\vct{r}\in\mathbb R^3$ and $t\in\mathbb R$ by
respectively the roton wavelength $a=\SI{3.26}{\angstrom}$~\cite{PitStr03,DonBar98} 
and a quantum typical time $t_0 = 2ma^2/\hbar =
\SI{1.34e-11}{s}$, where $m=\SI{6.65e-27}{kg}$ corresponds to the $^4\mathrm{He}$ atom
mass. Considering that the number of atoms is high in the condensate, we can
assume that the dynamics of $\psi$ is given by the GP equation, that reads in
its most general and dimensionless version
\begin{equation}\label{eq:GPNL}
i\frac{\partial \psi}{\partial t} = -\Delta \psi + (V\ast |\psi|^2)\psi - \mu \psi,
\end{equation}
where $V(\vct{x})$ is a smooth two-body interaction potential, assumed to be
spherically symmetrical (a function of the norm $x=|\vct{x}|$ only), $\ast$
stands for the convolution product, i.e. $(V\ast |\psi|^2)(\vct{x},t) = \int
V(\vct{x}-\vct{y})|\psi|^2(\vct{y},t)\,d^3y$, and $\mu = \int V(\vct{x}) \,
d^3x$ the chemical potential ensuring $|\psi|^2=1$ as a homogeneous solution.
As it is shown in Refs.~\cite{PomRic93,BerRob99,VilCas12}, the finite extension
of the interaction potential is crucial in obtaining a dispersion relation that
reproduces the roton minimum, as it is observed in neutron scattering
measurements performed in superfluid $^4\mathrm{He}$~\cite{DonBar98, fak2012}. 
This allows us to calibrate
the superfluid model that is proposed in Eq.~\ref{eq:GPNL} while computing the
dispersion relation, readily obtained as 
\begin{equation}\label{eq:DispRel}
\omega^2(\vct{k}) = |\vct{k}|^4+2|\vct{k}|^2\widehat{V}(\vct{k}),
\end{equation}
where $\widehat{V}(\vct{k})= \int  e^{i\vct{k}. \vct{x}}V(\vct{x})d^3 x$ is the
Fourier transform of the interaction potential, which depends only on
$k=|\vct{k}|$ if $V$ is taken isotropic. To get the dispersion relation (Eq.~\ref{eq:DispRel}), 
we use standard techniques consisting in linearizing Eq.~\ref{eq:GPNL} 
while looking for solution of the form $\psi=1+\varphi$, for
small $\varphi$, Fourier transforming the linear dynamics of $\varphi$ and its
conjugate, then looking for constraints on $\omega$ and $|\vct{k}|$ to avoid a
single trivial solution (see~\cite{PomRic93,BerRob99,VilCas12}). Choosing a
particular form for $\widehat{V}(\vct{k})$ allows then to compare the
so-obtained dispersion relation against experimental measurements.

In subsequent numerical simulations that we will present in the 
next sections, we
choose the following functional isotropic form for the interaction potential
\begin{equation}\label{eq:InterPot}
\widehat{V}(\vct{k}) = \left( \frac{c_s^2}{2}-v_1^2|\vct{k}|^2+v_2^4|\vct{k}|^4\right)\exp\left(-\frac{\left|\vct{k}\right|^2}{2k_0^2}\right),
\end{equation}
where $c_s$ corresponds to the sound velocity, i.e.\ the limit at small
wave-vector of $\omega^2(\vct{k})/|\vct{k}|$, and $(v_1,v_2,k_0)$ three free
parameters than could be obtained for example using a least-square fit
procedure given a experimental dispersion relation. We will not do that and
use instead, within our choice of units,  $c_s = 16$, $v_1=2.2635$,
$v_2=0.4408$ and $k_0=5.5970$. We now motivate our choice and compare against
experimental data.

We represent in Fig.~\ref{fig:DispRelStatSol}~(a) the dispersion relation of
liquid $^4\mathrm{He}$ at saturated vapor pressure provided in Ref.~\cite{DonBar98}
using a {solid black} line, and that exhibits indeed a roton minimum around
{$|\vct{k}|=2\pi$}. We superimpose there using a {red dashed-dotted line the dispersion relation obtained with the model proposed in Ref.~\cite{BerRob99}, which implications on the vortex density profile are discussed in section \ref{sec:SpaDisVor}. With a blue dashed line, we represent the dispersion relation} we get, using Eq.~\ref{eq:DispRel}, with an interaction potential (Fourier
transformed) given in Eq.~\ref{eq:InterPot} that we use in subsequent numerical
simulations, with the formerly defined free
parameters $(v_1,v_2,k_0)$ and the given sound velocity $c_s$. {The corresponding potential profiles in real space are represented in Fig.~\ref{fig:DispRelStatSol}~(b)}. We observe
quantitative differences between the experimental and our theoretical
dispersion relations. First, when expressed in physical units, we have chosen a
sound velocity of the order of \SI{354}{m/s}, which is  higher
than the observed value \SI{238}{m/s}. This explains why
experimental and theoretical curves deviate at vanishing $|\vct{k}|$.
Furthermore, we see that the theoretical curve reproduces the correct value of
the roton wavevector {$|\vct{k}|=2\pi$}, but not the value
of the of the minimum of angular frequency $\omega$ (nor consistently the value of the
\textit{maxon}, i.e.\ the local maximum of frequency occurring just before).
This choice is dictated by further numerical investigations in which we forbid
any crystallization phenomena, a natural tendency of this type of model 
(Eq.~\ref{eq:GPNL}) to evolve toward a periodical modulations of density
$\rho=|\psi|^2$, as it has been exploited to describe a possible
supersolid-state of matter~\cite{PomRic94,JosPom07}. {This tendency to crystallization is described in section \ref{sec:InterPot}.} At this stage, let us
state that the present approach, based on a scalar wave function with a
two-body non-local interaction as considered in Eq.~\ref{eq:GPNL} is unable to
describe the dynamics of $^4\mathrm{He}$ in a superfluid phase with a more realistic
dispersion relation. 

Indeed, let us show that such a choice for the interaction potential 
(Eq.~\ref{eq:InterPot}) allows axisymmetric stationary solutions, 
i.e.\ vortex-lines
with quantized circulation, as it has been widely studied for the local GP
equation~\cite{PitStr03}.

\section{A detour through the numerical estimation of axisymmetric stationary solutions}

A major success of scalar wave functional approaches, and its related dynamics
given by the GP equation~\cite{PitStr03}, lies in the existence of a stationary
solution (i.e.\ time independent) which is axisymmetric (say independent of the
$z$ coordinate and on the polar angle $\varphi$ in the $xy$-plane) and
exhibiting a $2\pi$ defect for the phase. More precisely, a solution of the
form $\psi(r,\varphi,z) = \sqrt{\rho(r)}e^{i\varphi}$, where we have introduced
the polar decomposition of the wave function in terms of amplitude and phase,
and the cylindrical coordinates $(r,\varphi,z)$ in which $x=r\cos (\varphi)$
and $y=r\sin (\varphi)$. To numerically estimate the shape of the density
distribution $\rho(r)$ as a function of the polar distance $r$, and more
generally to test the existence of such a axisymmetric solution, we numerically
solve the two-dimensional relaxation problem (corresponding to the propagation
of Eq.~\ref{eq:GPNL} in imaginary time with the $z$-independence as a
constraint)
\begin{equation}\label{eq:2Drelax}
\frac{\partial \psi}{\partial t} = \left(\frac{\partial^2 }{\partial x^2}+ \frac{\partial^2 }{\partial y^2}\right) \psi - (\widetilde{V}\ast |\psi|^2)\psi + \mu \psi,
\end{equation}
with initial condition $\psi(x,y,0) = e^{i\varphi(x,y)}$, $\varphi(x,y) =
\arctan(y/x)$ (the inverse tangent being suitably defined), and
$\widetilde{V}(x,y)=\int V(x,y,z) \, dz$. We solve this initial value problem
using periodic boundary conditions in order to efficiently compute 
linear operations in the Fourier space, and nonlinear ones in the physical
space. Doing so, in order to prevent from phase discontinuities, we use four
copies of this initial condition with appropriate phase distribution, and
evenly spaced, as it is explained and performed in Ref.~\cite{KopLev93}. In
units of the length scale $a$, we use as a mesh size $dx=1/16$, and depending
on the number of collocation points $N$ in each direction, we consider domains
of physical size $Ndx$. Using thus $N=512$ (vortices are sufficiently far
apart to neglect their interaction), we simulate a domain of physical size
$32\, a \approx 100\mathrm{\AA}$. Time propagation is performed using a
fourth-order Runge-Kutta explicit method with $dt={(dx)}^2/64$, corresponding to
\SI{8.2e-16}{s} in physical units. {This particular value for the time-step was obtained by starting from the numerical stability prediction for the heat equation $dt\leq dx^2/2$ \cite{PreTeu07}, and then decreasing $dt$ until all simulations were numerically stable.} To prevent from spurious generation of
unphysical small scales, we use as a dealiasing method  the $2/3$-rule, each
time we perform a multiplication in the physical space. In our case, since the
nonlinearity is of order three, we apply this rule {two} times at each time
step, which is enough to prevent the generation of unphysical small scales.

\begin{figure}[htbp]
\epsfig{file=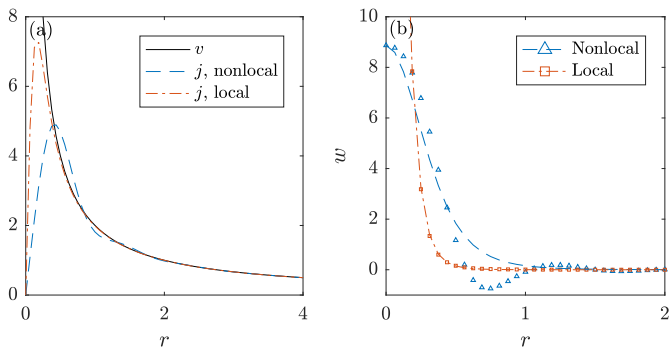,width=8.8cm}
\caption{(a) Radial distributions of the velocity field $\vct{v}$ and
probability current $\vct{j}$ of the axisymmetric stationary solutions of the
local (Eq.~\ref{eq:GPL}) and non-local (Eq.~\ref{eq:GPNL}) GP equations. (b)
Radial distributions of the associated pseudo-vorticity fields (Eq.~\ref{eq:PsVor}), 
and their comparison with a schematic fit provided in 
Eq.~\ref{eq:ModMonDecrease}.\label{fig:RadialDistrib}}
\end{figure}

Starting from our initial condition, we observe the convergence of Eq.~\ref{eq:2Drelax} 
toward a time-independent solution that we can consider as a
stationary solution of the non-local GP equation (Eq.~\ref{eq:GPNL}) itself, and
we represent it in Fig.~\ref{fig:DispRelStatSol}~{(c)}. {We see that indeed the
solution is symmetrical around the axis of the vortex, where
the density tends to zero, as it is expected at the core of quantum vortices}. We
furthermore see the existence of additional density oscillations around the
vortex core, which are themselves expected once a roton minimum is included in
the picture. This is a well-known 
phenomenon~\cite{Reg72,Dal92,BerRob99,RosVit12,VilCas12,MolGal16}, for which density
oscillations are prescribed by the characteristic shape and size of the roton
minimum. We represent in the inset of Fig.~\ref{fig:DispRelStatSol}~(b) a
comparison of the density profile away from the {vortex core}
obtained in axisymmetric stationary solutions of the local (given in Eq.~\ref{eq:GPL}, 
curve is in black) and non-local GP equations (our present
numerical estimation) where we see more clearly that the presence of the roton
minimum leads to periodical modulation of density, governed by the roton
characteristics, and can be seen as precursors of the crystallization
phenomenon \cite{PomRic94,JosPom07}. 

\section{Further comments on the model of the interaction potential} 
\label{sec:InterPot}

We represent (red {dashed-dotted} line) in Fig.~\ref{fig:DispRelStatSol}~(a) the model of
the non-local interaction potential $V$ chosen in Ref.~\cite{BerRob99}, which
also includes an additional quintic term in the overall dynamics. It does
reproduce with great accuracy the experimental dispersion relation of 
Ref.~\cite{DonBar98}. Nonetheless, when looking for a possible axisymmetric
stationary solution, in the spirit of the relaxation problem posed in Eq.~\ref{eq:2Drelax}, 
we end up eventually with a time-independent solution
displayed in Fig.~\ref{fig:DispRelStatSol}~{(d)}. We see that in this case, the
invariance around the axis of the {vortex} is broken by the appearance of
additional periodic modulation of density that form a hexagonal structure. This
crystallization phenomenon was already observed in early numerical simulations
of the non-local GP equation~\cite{Ber99}, where it is associated to a
phenomenon of mass concentration associated to negative values of the
interaction potential. Remark also that we could have used also another set of
parameters $(c_s,v_1,v_2,k_0)$ in order to be much closer to the experimental
dispersion curve, and without including an additional quintic term. Similarly
{to} the model proposed in Ref.~\cite{BerRob99}, the corresponding
stationary vortex solution would also exhibit such a crystallization
phenomenon (data not shown). {We performed similar simulations using domains of larger sizes (up to $2048^2$ collocation points using the same $dx$, and not displayed here for the sake of clarity) and observe a similar crystallization
phenomenon with the same period. As we will comment while evoking the notion of supersolidity as it is discussed in Refs. \cite{PomRic94,JosPom07}, the period of such a precursor of crystallization  is governed by the finite extension of the interaction potential, and is thus not influenced by the macroscopic scale of the domain. Let us furthermore mention that these stationary solutions are obtained while solving a 2D relaxation problem, which includes as an additional constraint the translational invariance along the vortex axis (Eq. \ref{eq:2Drelax} is readily obtained from Eq. \ref{eq:GPNL} assuming independence on the spatial variable $z$). Thus, these density oscillations should be furthermore seen as being independent of the $z$-direction, which is physically barely acceptable. For these reasons, we decide to look for a set of parameters $(c_s,v_1,v_2,k_0)$ that would allow a non-crystallized stationary vortex solution.}

In the light of more recent studies~\cite{PomRic94,JosPom07} concerning the
natural evolution toward the state of supersolidity when a roton minimum is
included in the picture, we are left with concluding that a GP type of
evolution, where enters a simple non-local two-body interaction (with a possible
additional quintic term), is unable to describe in a proper manner superfluid
$^4\mathrm{He}$ if we follow with great accuracy the experimental dispersion curve of
Ref.~\cite{DonBar98}. {Let us note that, to our knowledge, there is no experimental evidence of the very microscopic structure of vortices in superfluid helium at the roton scale. As a consequence, we cannot affirm that such a crystalline structure is impossible in $^4\mathrm{He}$. However, at a macroscopic scale, non-crystallized vortices are well supported by experimental evidence in superfluid Helium \cite{YarGor79} and in atomic Bose Einstein Condensates (BEC) \cite{WilNew15}. In other words, we ask the model of superfluid we consider and given in Eq. \ref{eq:GPNL} to exhibit a non-crystallized vortex (i.e. axisymmetric) as a possible stationary solution. To do so, we are led to use an interaction potential $V$ which is unable to reproduce with great accuracy the experimental dispersion curve.} Note nonetheless that calibrating our model, which
involves a nonlinearity in the evolution (Eq.~\ref{eq:GPNL}), using the
dispersion relation, that is a prediction obtained through a linearization
procedure, is difficult to control. It would be of great interest to develop a
new type of dynamics that would include both a correct description of the
experimental dispersion relation, and the existence of stationary axisymmetric
solutions representing in a proper way quantum vortices. We leave this aspect
for future investigations, and perform subsequent three dimensional numerical
simulations with the aforementioned model for the interaction potential 
(Eq.~\ref{eq:DispRel}), that prevents the formation of these hexagonal periodic
modulations of density.

\section{Spatial distribution of the vortex solution} 
\label{sec:SpaDisVor}

Let us now investigate the very radial distribution of various kinematic
quantities entering in the hydrodynamical interpretation of the non-local GP
equation (Eq.~\ref{eq:GPNL}), as it is given by the Madelung 
transformation~\cite{PitStr03}. In this approach, we associate  the gradient of the phase of
$\psi$ to a velocity field $\textbf{v}$ and $|\psi|^2$ to a local density field
$\rho$. Key kinematic quantities are thus density $\rho = |\psi|^2$ and
probability current $\vct{j}= -i(\psi^* \vct{\nabla}\psi-\psi
\vct{\nabla}\psi^*) = \rho \vct{v}$ that are governed by conservation equations
that read~\cite{BerRob99,NorAbi97,PitStr03}
\begin{equation}
\frac{\partial \rho}{\partial t} + \vct{\nabla}.\vct{j}=0,
\end{equation}
that can be interpreted as a continuity equation, and considering a component $j_i$ of the vector $\vct{j}$, we have
\begin{equation}
\frac{\partial j_i}{\partial t} + \partial _j\Pi_{ij}=0,
\end{equation}
where $\Pi_{ij}$ is the momentum tensor,
\begin{equation}
\Pi_{ij} = \partial_i \psi \partial_j \psi^* -\psi\partial^2_{ij}\psi^*+\mbox{c.c.} + \frac{1}{2}\rho (V\ast \rho)\delta_{ij},
\end{equation}
where $\mbox{c.c.}$ stands for the complex
conjugate, and $\delta_{ij}$ the Kronecker symbol. Similarly, we could derive
the time evolution of the velocity field $\vct{v}$, which corresponds to a
compressible, irrotational and barotropic fluid with an additional quantum
pressure term, of density corresponding to $|\psi|^2$ (see for instance 
Refs.~\cite{BerRob99,Ber99}). It is well known that the velocity field diverges in
the vicinity of a defect of the phase, so we will in the next sections work with the
current vector $\vct{j}$, that is eventually, as we are going to see, a bounded
vector.

We display in Fig.~\ref{fig:RadialDistrib}~(a) the radial profile of velocity
$\vct{v}$ and probability current $\vct{j}$ of the axisymmetric stationary
solution represented in Fig.~\ref{fig:DispRelStatSol}~(d). Let us recall that,
using cylindrical coordinates centered on the vortex, the velocity field
$\vct{v} = 1/r \, \vct{e}_\theta$ and the current $\vct{j}=\rho(r)/r \,
\vct{e}_\theta$ are known. Remark that $|\vct{v}|=1/r$, as we mentioned,
diverges in the vicinity of the vortex line, whereas $\vct{j}$ is a bounded
vector since $\rho(r)$ tends to 0 as $r^2$~\cite{PitStr03} in the vicinity of
the origin (so that $\vct{j}$ vanishes itself at the origin). This is what is
indeed obtained and displayed in Fig.~\ref{fig:RadialDistrib}~(a). We
furthermore superimpose the radial distribution of $\vct{j}$ that is obtained
from the local, i.e.\ standard, version of the GP equation, that we define
precisely later (see Eq.~\ref{eq:GPL}). Once again, we see that the current
follows a non monotonic radial behavior. Compared to what is obtained in the
non-local GP equation (Eq.~\ref{eq:GPNL}), we can see that the maximum of
current is obtained at a similar distance from the axis of symmetry in both
models, but its value is higher in the non-local version of the dynamics.

Of great interest also from a dynamical point of view is the radial
distribution of the pseudo-vorticity
\begin{equation} \label{eq:PsVor}
\vct{w} = \vct{\nabla}\wedge \vct{j},
\end{equation}
which quantifies the rotational motions of this compressible fluid. In the following, we note $w = \left|\vct{w}\right|$ the amplitude of pseudo-vorticity. For a single vortex line along the $z$ axis, $\vct{w}$ is aligned with $\vct{e}_z$, and its amplitude only depends on the radial distance $r$ to the axis, and we have $w(r) = (1/r)d\rho(r)/dr$. Recall that
vorticity itself, i.e.\ the curl of the velocity field $\vct{v}$, is of
distributional nature, and  vanishes everywhere except at the {very center of the vortex core} where it diverges. Instead, as we can see in 
Fig.~\ref{fig:RadialDistrib}~(b), $\vct{w}$ as defined in Eq.~\ref{eq:PsVor}, 
is a bounded quantity.

In the local case (red symbols), we can even see that the
radial distribution of pseudo-vorticity, as far as the axisymmetric solution is
concerned, follows a monotonic decrease away from its axis of symmetry. As we
see in section \ref{sec:NumInv}, in particular to interpreted some quantities entering in
a statistical description of the flow, it is interesting to design a model for
this radial behavior. For this purpose, we propose the following decreasing
function
\begin{equation}\label{eq:ModMonDecrease}
w(r) = \frac{w_0}{{\left[ 1+{\left(\frac{r}{r_0}\right)}^2\right]}^\alpha},
\end{equation}
where $w_0$ is the value of pseudo-vorticity on the axis of symmetry, and
$(r_0,\alpha)$ two free parameters describing the shape of pseudo-vorticity
radial distribution. Starting with the radial distribution of $\vct{w}$
obtained from the local GP equation (and presented later in Eq.~\ref{eq:GPL}),
the fit (Eq.~\ref{eq:ModMonDecrease}) reproduces in a fairly good way the
observed decrease using $r_0 = 0.22$ (in units of $a$) and $\alpha=3.125$. In a
non-local context setting, as given by Eq.~\ref{eq:GPNL}, such a fit reproduces
accurately the decrease with $r_0 = 0.62$ and same $\alpha$, but
obviously fails at reproducing the non monotonic behaviors associated to the
additional oscillations associated to the roton minimum. As we will see in section \ref{sec:NumInv}, even
if some aspects are not reproduced by such a schematic fit 
(Eq.~\ref{eq:ModMonDecrease}), it will be very useful to interpret subsequent
statistical quantities that we will observe in the next sections. 

{Let us also remark that the exponent $\alpha\approx 3.125$ is not that close to the value $2$, a typical value that is expected while considering the boundary value problem related to the spatial distribution of vortex density. Indeed, far from the vortex origin, and neglecting second order variations, density should reach the uniform solution as the square of the inverse of the distance from the vortex \cite{PitStr03}. Using then $w(r) = (1/r)d\rho(r)/dr$, we easy get that pseudo-vorticity amplitude should tend to zero as $1/r^4$, corresponding thus  to $\alpha=2$. In our case, we are looking at the behavior of pseudo-vorticity over a range that cannot be considered as being far from the origin of the vortex, namely over say one atomic distance (see the obtained values for the parameter $r_0$). Moreover, our simulation domain is finite, and copies, that are necessary to prevent from phase discontinuities in this periodical set-up, might interact in a subtle way such that the value $\alpha\approx 3.125$ reproduces in a better way the decrease of the pseudo-vorticity that we are observing. This may explain why we are not observing a $1/r^4$-decrease of pseudo-vorticity.}

\begin{figure*}[htbp]
\begin{center}
\epsfig{file=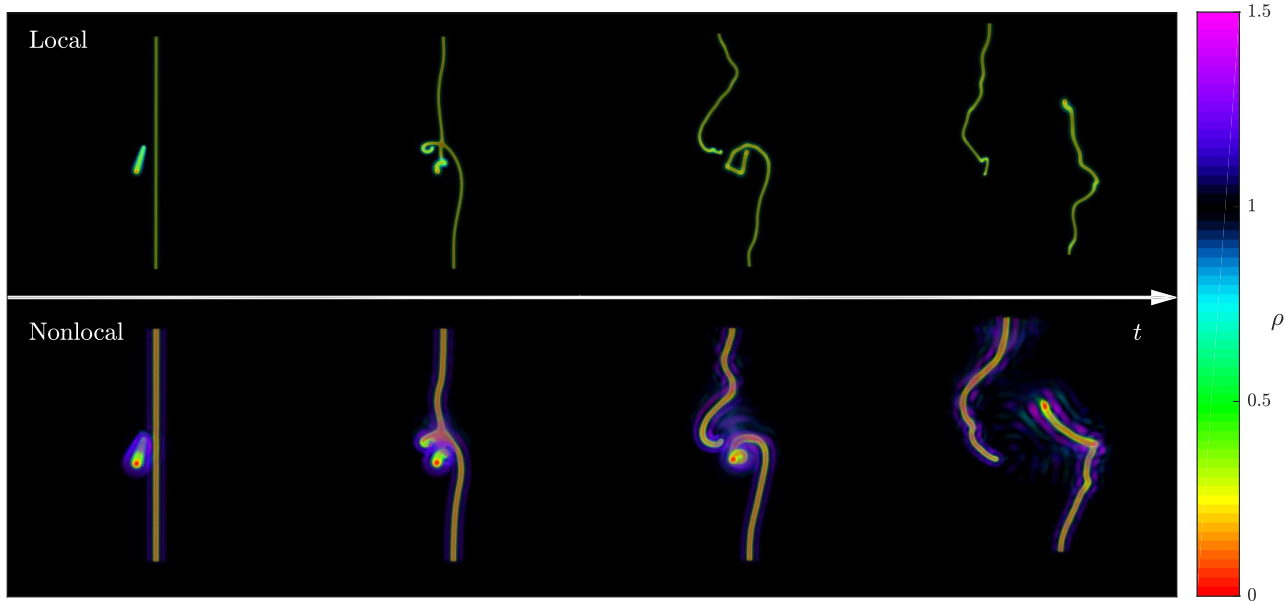,width=17cm}
\end{center}
\caption{3D visualization of the superfluid density during reconnection, in the
local (top) and non-local (bottom) models. A density threshold is applied for
clarity, so that the bulk fluid density $\rho\sim1$ is transparent. Four
snapshots are represented along time, from left to right respectively: the
initial condition $t=0$, the reconnection time $t=t_{\text{rec}}$, twice the
reconnection time $t=2t_{rec}$, and the end time of the simulation $t =
t_{\text{end}}\approx 6t_0$. \label{fig:RepreReconn}}
\end{figure*}

\section{Numerical investigation of reconnection of a set of initially perpendicular vortices}
\label{sec:NumInv}

Let us now investigate the dynamics of vortex reconnection in the presence of a
roton minimum, and thus with a non-local interaction potential (Eq.~\ref{eq:GPNL}), 
as it was studied in a local GP equation in Ref.~\cite{KopLev93}. 
To do so, we prepare an axisymmetric stationary solution as it
is described in the former paragraph, and represented in 
Fig.~\ref{fig:DispRelStatSol}~(d), properly extended to three dimensions and take as
an initial condition the product of two such wave functions, one being in the
center of the domain and the other one being shifted from the center by one atomic
distance $a$ and rotated such that we get initially two perpendicular
vortices, in a very similar way as in Ref.~\cite{KopLev93}. One of the purpose
of the present article is to study and quantify the differences in the
evolution of this set of two perpendicular vortices with and without the roton
minimum. Thus, in addition to three-dimensional numerical simulations of the
non-local GP equation (Eq.~\ref{eq:GPNL}), as we have already discussed in particular
in Fig.~\ref{fig:RadialDistrib}, we perform such a simulation with the standard
(local) form of the GP equation. In our system of units, we thus consider also
the following dynamics
\begin{equation}\label{eq:GPL}
i\frac{\partial \psi}{\partial t} = -\Delta \psi + g\left(|\psi|^2-1\right)\psi,
\end{equation}
{where we have implicitly chosen $\mu=g$ as a chemical potential to ensure that the stationary solution is defined by $\psi=1$. In both models given in Eqs. \ref{eq:GPNL} and \ref{eq:GPL}, the speed of sound $c_s=\sqrt{2\mu}$ and the typical vortex core extension (i.e. the healing length) $\xi=1/\sqrt{\mu}$ depend directly on the value of the chemical potential $\mu$. In the nonlocal case Eqs. \ref{eq:GPNL}, we have chosen $\mu= \widehat{V}(0)=c_s^2/2=16$ (Eq. \ref{eq:InterPot}), a value that was shown to prevent from crystallization events. This leads us to choose for the coupling constant entering in Eq. \ref{eq:GPL} the value $g=128$ in order to work with same $c_s$ and $\xi$.} {Similarly} as it is
done while considering the relaxation problem of Eq.~\ref{eq:2Drelax}, we solve
Eqs.~\ref{eq:GPNL} and~\ref{eq:GPL} in a periodic {domain} using a
pseudo-spectral method, with same $dx$, $dt$ and dealiasing rule, over
$N^3=512^3$ collocation points. Results of both simulations are displayed in
Fig.~\ref{fig:RepreReconn}, using the visualization software Vapor~\cite{ClyMin07}. 
Once again, only one eighth of the
computational domain is displayed, but we keep in mind that copies remain to warrant a
continuous distribution of the phase of the wave function.

We have displayed the evolution of this set of two initially perpendicular
vortices at four different time: (i) the initial time $t=0$, (ii) at the time
of reconnection $t=t_{\text{rec}}$, (iii) after the reconnection at
$t=2t_{\text{rec}}$ 
and (iv) some time after the reconnection at $t=10t_{\text{rec}}$. 
{The time of reconnection are similar in both dynamics, namely, $t_{\text{rec}}$ is 1.16 in the local formulation,
and 1.08 in the non-local one, resp.\ \SI{1.55e-11}{s} and \SI{1.44e-11}{s}
in physical units.
Since we have chosen the parameters of the dynamics such that the speed of sound is the same in both the local and non local formulations, it is not surprising to observe similar reconnection times $t_{\text{rec}}$.} However, it is very surprising at this stage that overall the
phenomenon of reconnection, from a spatial point of view, looks very similar in
the local and non-local formulation. Indeed, the spatial distributions of
density $|\psi|^2$ at the final stage $t_{\text{end}}$ that we consider in both cases
share striking similarities. We can conclude that the internal (i.e.\ 
microscopic) structure of the vortices, strongly influenced by the presence of
the roton minimum, has nonetheless little influence on the overall global
evolution at larger scales than $a$. {One could argue that in the very core of the vortex the nonlinearity of the GP model has little influence on the reconnection dynamics. However in the presence of a roton minimum the internal core is always surrounded by strong density fluctuations, and we are solving a nonlinear problem that could have been highly sensitive to those density structures.} From a local point of view,
paying attention to the precise values of the densities across time and space,
we can see that the non-local GP allows locally high values (of the order of
$\rho\sim 1.5$), which we recall do not seem to have implications on
the global dynamics.

\begin{figure*}[t]
\begin{center}
\epsfig{file=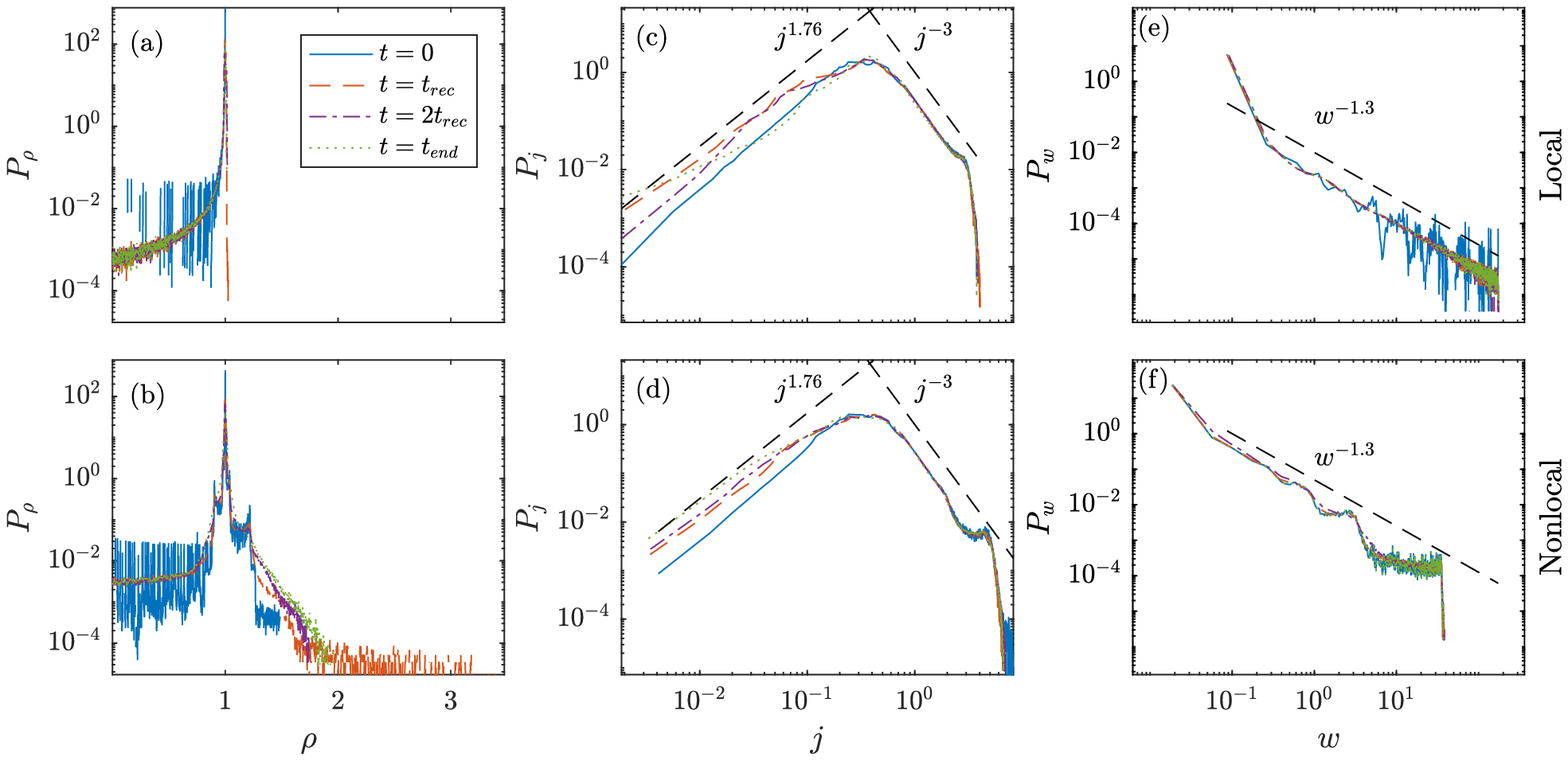,width=17cm}
\end{center}
\caption{\label{fig:PDFFields} Plots of the probability density functions (PDF)
of various hydrodynamical quantities, in the local (left) and non-local (right)
models. (a,b): PDF of the superfluid densities $|\psi|^2$. High density events
$\rho>1$ are well represented in the non-local case (b), whereas they are absent
in the local model (a). (c,d): PDF of the probability current norm $j =
\left|\vct{j}\right|$ (see text for a precise definition). A $j^{-3}$ behavior
is observed on the right tails of both models and comes from the midrange decay
of $j$. A $j^{1.76}$ scaling is observable on both left tails. (e,f): PDF of
the pseudo-vorticity norm $w = \left|\vct{w}\right|$ (Eq.~\ref{eq:PsVor}). The
observed power-law behaviors $P_w\sim w^{-1.4}$ are super-imposed in both the
local and non-local cases.}
\end{figure*}

To quantify more precisely the generation of strong fluctuations of the
superfluid density $\rho=|\psi|^2$, we compute the probability density function
(PDF) of the field of densities obtained in the simulation domain for both the
local and non-local cases, at the four times considered in Fig.~\ref{fig:RepreReconn}, 
and we display our results in Figs.~\ref{fig:PDFFields}~(a) and (b). We observe that
along the phenomenon of
reconnection, densities higher that the uniform one are
forbidden in the local case (Eq.~\ref{eq:GPL}). On the contrary, in the
non-local approach, in which local densities higher than one are initially
present due to the roton minimum, the dynamics may develop local mass
concentration exceeding three times the value of the uniform density.

In order to interpret subsequent PDFs that we are going to estimate, let us
first focus on the simple axisymmetric stationary solution (i.e.\ the vortex
line) that we presented in Fig.~\ref{fig:DispRelStatSol}~(d). Consider then any
physical quantity of interest $F$, and its respective PDF $\mathcal P_F$, i.e.\ 
the histogram of the values $g(x,y,z)$ taken by the quantity $F$ in the domain
$\mathcal V$ (of volume $|\mathcal V|$). The PDF can be written as the
following empirical average
\begin{equation}\label{eq:EmpInterpPDF}
\mathcal P_{F}(f)=\frac{1}{|\mathcal V|}\int_{\mathcal V} \delta \left( f-g(x,y,z)\right)dxdydz,
\end{equation}
where $\delta$ denotes the Dirac function. For a single vortex, in a
cylindrical volume $\mathcal V$ of radius $R$ and of finite height, the PDF of
$\vct{v}$ can be computed {exactly}, inserting $g(r,\theta,z)=1/r$ in
the empirical interpretation of the PDF (Eq.~\ref{eq:EmpInterpPDF}) and
performing a proper change of variable, we get $\mathcal
P_{|\vct{v}|}(|\vct{v}|) = 2R^{-2}|\vct{v}|^{-3}$ for $|\vct{v}| \ge R^{-1}$
(and 0 for $|\vct{v}|<R^{-1}$), showing that the tail is governed by the
divergence of velocity in the vicinity of the vortex, as it was noticed in 
Ref.~\cite{PaoFis08,ProNaz12}. Note that similar power-law behavior $\sim|\vct{v}|^{-3}$ have
been observed in simulations of the local GP equation (Eq.~\ref{eq:GPL}) as
detailed in Ref.~\cite{WhiBar10}. This power-law behavior of the tail of the
PDF of the norm of velocity is also observed for the case of two perpendicular
vortices, as we consider to initiate our numerical simulations. We have checked
in our simulation that this is also the case for the initial condition we are
using, for both the local and non-local case, all long the reconnection process
(data not shown).

To this regard, as we have already seen, the probability current field
$\vct{j}$ remains bounded in the presence of a vortex line, and thus appears to
be a good candidate in order to quantify whether high values of density, as
they are observed in particular in the non-local case (Fig.~\ref{fig:PDFFields}~(b)), 
are associated to high values of the current
$\vct{j}$. We display in Figs.~\ref{fig:PDFFields}~(c) and (d) the histograms of
the values taken by the norm of the current vector $\vct{j}$ in the
computational domain. We see in both local and non-local cases, that the PDF of
$|\vct{j}|$ exhibits similar tendencies, such as (i) a linear trend for
$|\vct{j}|\ll 1$, and
(ii) a $|\vct{j}|^{-3}$ power-law behavior in a domain of finite extension,
reminiscent of the expected power law of the velocity field PDF, as explained
formerly. The power law behavior at small $|\vct{j}|$ should be a consequence
of the interaction of vortices and their images, and the finiteness of the
extension of the computational domain (and the respective periodical boundary
condition). Even if higher values of the current field $|\vct{j}|$ are indeed
observed in the non-local case (Fig.~\ref{fig:PDFFields}~(d)), they do not exceed
values already observed in the initial condition at $t=0$. We are lead to the
conclusion that the existence of a roton minimum has little influence on the
overall shape of the PDFs of $|\vct{j}|$. 

Developing on these ideas, we perform a similar estimation of the histogram of
the pseudo-vorticity $\vct{w}$ (Eq.~\ref{eq:PsVor}), in order to quantify
possible creation of \textit{small scales}, as it happens in the presence of a
direct cascade mechanism, which is at the heart of the phenomenology of
Kolmogorov regarding three-dimensional classical (i.e.\ governed by the viscous
Navier-Stokes equation) turbulence~\cite{Fri95}. Indeed, recall that in
classical turbulence, velocity PDF is close to a Gaussian function, whereas PDF
of gradients, and in particular vorticity, is found highly non Gaussian. For a
vortex line along the $z$-axis, we know well that pseudo-vorticity is expected
to be  a bounded vector, taking significant values only for $r$ of the order
and smaller than $a$. Using thus the schematic distribution provided in 
Eq.~\ref{eq:ModMonDecrease}, it is easy to get, from Eq.~\ref{eq:EmpInterpPDF},
that we expect $\mathcal P_{|\vct{w}|}(|\vct{w}|)\propto
|\vct{w}|^{-1-\alpha^{-1}}$. This assumption on the radial profile of
pseudo-vorticity allows to reproduce the present observed histograms of the
values taken by  $|\vct{w}|$ in our simulation domain at any time of the
reconnection process, as it is displayed in Figs.~\ref{fig:PDFFields}~(e) and
(f). {According to this model, using $\alpha \approx 3.125$ for both the local and
non-local cases, we expect a power law decrease of the PDF
with an exponent $\approx 1.3$, as it is presently observed. As we mentioned formerly, for a single stationary vortex, the large $r$ asymptotic for the pseudo-vorticity gives $|\vct{w}|\propto r^{-4}$, which suggests $\alpha=2$. It gives a different power law exponent for the PDF of $|\vct{w}|$, namely $|\vct{w}|^{-1.5}$ instead of our observed $|\vct{w}|^{-1.3}$ behavior. Once again, this difference could be explained by the fact that the PDF that is measured here is for two vortices during reconnection in a finite periodic domain, contrary to the large $r$ asymptotic that is derived for a single, straight vortex in infinite space.}

{We would like now to comment briefly on the amount of sound emitted following the reconnection event. Such a study has been carried out in the literature using the local formulation of the GP equation (Eq. \ref{eq:GPL}) \cite{NorAbi97,DilMin17}, that we repeat here for the nonlocal version. Decomposing the conserved total energy as the sum of various components, we identify the part associated to kinetic energy as being the average (over space and time, the time integration starting at $t_{rec}$ until the end of the simulation) of the norm square of the vector $\sqrt{\rho}\vct{v}$, a vector field being itself decomposed into a divergence-free part (i.e. incompressible) and a compressible part, a decomposition easily performed in the Fourier space. Doing so, we obtain, in average, the incompressible $E_{\text{kin}}^{\text{i}}$ and compressible $E_{\text{kin}}^{\text{c}}$ kinetic energies. In present numerical simulations, we obtain for the ratio $E_{\text{kin}}^{\text{c}}/E_{\text{kin}}^{\text{i}}$ of these energies a value of $0.045$ in the local case, and $0.073$ for the non local version. Although more sound is emited in the non local case, overall it shows that the most part of the kinetic energy is held by the incompressible motions.}

This being said,
even if there are some differences implied by the existence of the roton
minimum, we are led to the conclusion that there is no creation of small
scales, i.e.\ no creation of high value of pseudo-vorticity, even in the
presence of a model taking into account a realistic picture of the core of
vortices. We will come back to this point in the conclusion.

\begin{figure}[htbp]
\begin{center}
\epsfig{file=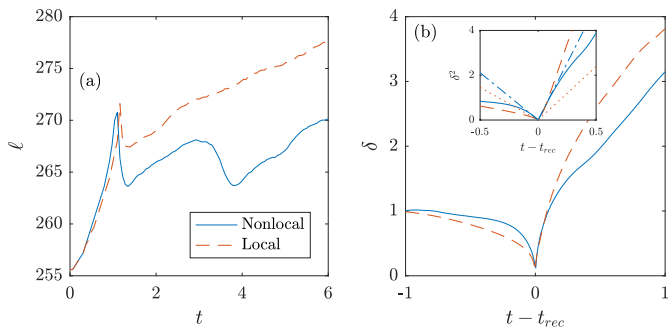,width=8.8cm}
\end{center}
\caption{\label{fig:LengthApp} (a): Total length of the set of two vortices
(see text), in the local (red) and non-local (blue) models. (b): Distance $\delta$
between the two vortices, in the local (red) and non-local (blue). Inset: squared distance $\delta^2$ between the two vortices for both models and their respective linear fit (dashed and dotted-dashed lines). The slopes of the linear fits before and after reconnection are given in the text.}
\end{figure}

\section{Tracking vortices}

Let us now explore some other aspects of the vortex reconnection process, as
those related to the evolution of vortices as individual objects. Such kind of
studies rely on the tracking of vortex cores, i.e.\ regions of space where
density vanishes. This can be done using algorithms that seek zeros in planes,
in order to extract lines in three-dimensional space with the help of the
pseudo-vorticity, as it is proposed in recent literature~\cite{VilKrs16}. In
this section, we revisit what has been done in this context for the local GP
equation~\cite{RorSki16,VilPro17,ZucBag12}, and in experiments~\cite{FonMei14}, and
compare with what is obtained in the non-local case. In few words, this
numerical tracking algorithm provides the position vector $\vct{X}(s,t)$ at
each time $t$ and parameterized by the length $s$  measured along the filament
made of the points that do not hold a density. From there, we can define the
Frenet-Serret frame of reference, given by the orthonormal set of vectors
$(\vct{T},\vct{N},\vct{B})$, where $\vct{T}=\partial \vct{X}/\partial s$ is the
tangential vector, $\vct{N} = (\partial \vct{T}/\partial s)/|\partial
\vct{T}/\partial s|$ the normal vector, and $\vct{B}=\vct{T}\wedge \vct{N}$ the
binormal vector. In a equivalent way, we could have introduced in the
definition of this frame the curvature $\kappa(s,t) = |\partial
\vct{T}/\partial s|$ and torsion $\tau(s,t)=-\vct{N}\cdot \partial
\vct{B}/\partial s $, where $\cdot$ stands for the scalar product 
(see Ref.~\cite{Has72} for instance).

In subsequent developments, we will analyze the vortex reconnection phenomenon
in the Frenet-Serret frame, and compare with the well-known schematic model
given by the local induction approximation (LIA). LIA has been studied for a
long time in various aspects of fluid dynamics, and can be derived in a
systematic way from the Navier-Stokes equation~\cite{CalTin78} assuming that
the vortex core size is small compared to some characteristic curvature. From a
dynamical point of view, this approximation implies that the time variation of
the position $\vct{X}(s,t)$  for some time independent parameterization $s$ has
only a contribution along the binormal vector, namely $\dot{ \vct{X}}=\partial
\vct{X}(s,t)/\partial t = G\kappa(s,t) \vct{B}(s,t)$, where $G$ is a constant
that diverges {logarithmically} with the vortex core size, and $\kappa
(s,t)$ the local curvature. The LIA predicts in a consistent way that indeed
the length of such a vortex is conserved, and that solitary waves (i.e.\ 
solitons) can propagate at a constant speed~\cite{Has72}. Let us then compare
these predictions against the dynamics of the reconnection process that we
observe in our numerical estimations of the local and non-local GP equations.

We display in Fig.~\ref{fig:LengthApp}~(a) the length of the system made up of
the two vortices, as they are displayed in Fig.~\ref{fig:RepreReconn}. We
furthermore include in the estimation of their lengths their copies in the
whole simulation domain. Remark that the aforementioned tracking algorithm
provides in a straightforward manner their lengths. We see that in both local
and non-local cases, before reconnection, vortices undergo stretching, that make
their length increases of a small amount (of order $5\%$) before decreasing. It
follows then, after reconnection, a monotonic increase for the local case, and
a more complex evolution for the non-local case.  As we claimed, LIA predicts
that the length is a constant a motion. Indeed, defining the length of a vortex
as $\ell(t) = \int |\partial \vct{X}(s,t)/\partial s |ds$, we get from a
general point of view (for any parameterization $s$) that $d\ell(t)/dt = \int
\partial \dot{ \vct{X}}/\partial s \cdot \vct{T}ds$, showing that a non
vanishing component of the induced velocity $\dot{ \vct{X}}$ along the normal
vector $\vct{N}$ may contribute to a variation of the length $\ell$, which is
not the case in the LIA (only a component along the binormal is considered).
Such arguments on vortex length have been rigorously studied in~\cite{KleMaj91}. 
As we can see in Fig.~\ref{fig:LengthApp}~(a), the length of
vortices depends on time, a feature that is not allowed in the LIA, although
only $5\%$ of the length undergo changes.

To carry on the description of the dynamical features of reconnection, we
compute the distance separating vortices before and after reconnection, which
is defined at each time as the minimum distance between two points on the two
vortex lines. Such a numerical study has been performed {systematically}
for the local GP equation for various initial conditions~\cite{VilPro17}, and
it was found that this distance behaves as $|t-t_{\text{rec}}|^{1/2}$,
both before and after reconnection. The square-root behavior can be understood
using a linear approach, justified close to the vortex core (where density
vanishes)~\cite{NazWes03}, or from a dimensional point of view (see for
instance Ref.~\cite{HorBre12}). The proportionality constant to this
square-root law was found to depend on initial conditions, whether, as an
example, vortices are taken perpendicular of anti-parallel. We represent in 
Fig.~\ref{fig:LengthApp}~(b) the time evolution of this distance $\delta$ before and after
reconnection in our present numerical simulations (see also a representation of  $\delta^2$ in the inset). For both the local and
non-local cases, we reproduce the square-root behavior close to the reconnection time $t_{\text{rec}}$, but with slightly different proportionality constants. This numerical estimation shows that the roton minimum has here some influence, in particular we see that approaching and separation distances have different time evolution. The numerical values we observe for the proportionality constant in the local case (resp. nonlocal) are significantly different from the values observed in ref. \cite{VilPro17}. We find $0.23$ (before reconnection) and $0.38$ (after reconnection) as far the local case is considered. Concerning the nonlocal case, we find $0.33$ (before $t_{\text{rec}}$) and $0.81$ (after $t_{\text{rec}}$). We recall that it was found in Ref. \cite{VilPro17}, for the local case, the corresponding values $0.55$ and $0.63$. Remark that, in this study, the two orthogonal vortices are initially separated by an atomic distance $a$, whereas this initial distance was chosen to be six healing lengths (which is of the same order of $a$), thus a factor of order 6 for the chosen initial separations. This could explain once again the strong dependence of this multiplicative constant on initial conditions, and the differences in between the present numerical study and the one proposed in Ref. \cite{VilPro17}. {However it is worth noticing that even if we observe slightly different values of the proportionnality constants in the local model, in both our case and the observation of Ref.~\cite{VilPro17} the value of the constant after the reconnection is always greater than the value of the constant before reconnection.}

\begin{figure*}[t]
\begin{center}
\epsfig{file=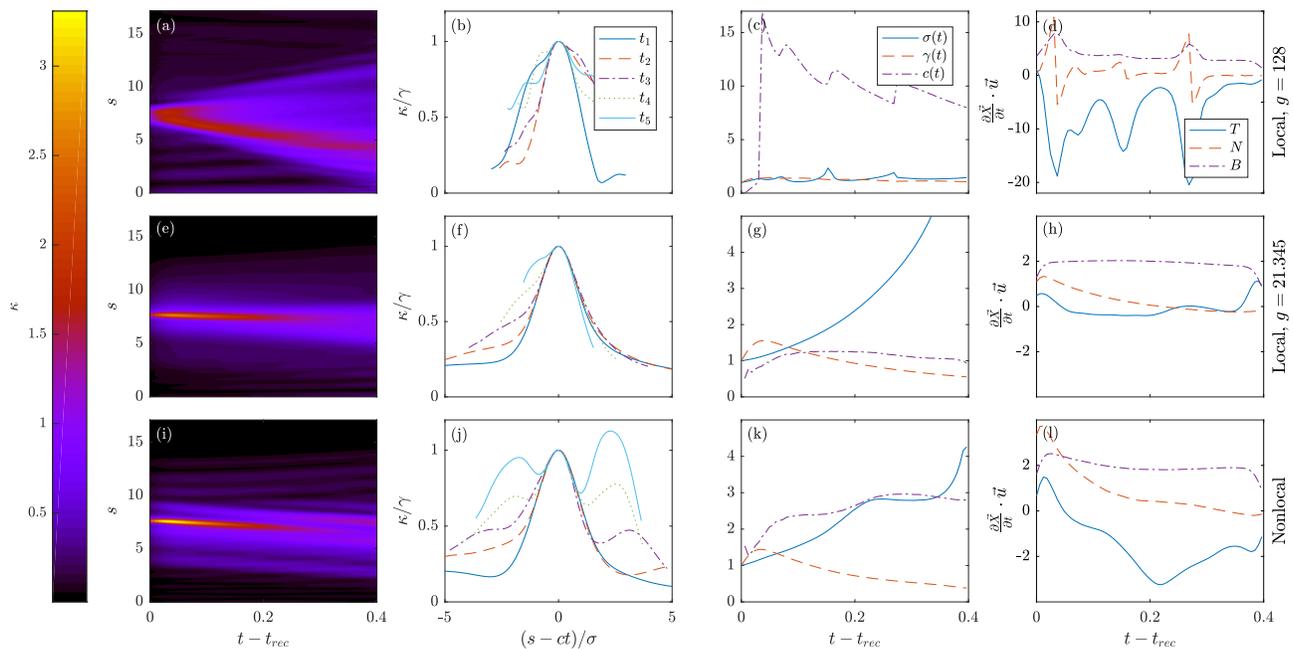,width=17cm}
\end{center}
\caption{\label{fig:Soliton} {(a,e,i): space-time maps of the local curvature of a
vortex line $\kappa(s,t)$, in the local case for $g=128$ (in (a)) and $g=21.345$ (in (e)),  and for the non-local model (in (i)). The
position of the principal curvature maximum is tracked along time. (b,f,j):
corresponding curvature profiles of the tracked maximum at five regularly spaced times during
its propagation. Each profile is scaled back onto the initial condition at
reconnection. Snapshot times are $t_1=t_{\text{rec}}$, $t_2=t_{\text{rec}}+0.09$, $t_3=t_{\text{rec}}+0.17$, $t_4=t_{\text{rec}}+0.26$ and $t_5=t_{\text{rec}}+0.34$. We observe the growth of
secondary curvature maxima in the non-local case (see (j)). (c,g,k): time evolution of
the $\sigma$, $\gamma$, $c$ scaling parameters, which represent the width, height and
velocity of the curvature {wave packet} respectively (see Eq.~\ref{eq:ABcSoliton}).
(d,h,l): projections of the velocity vector of the local maximum of curvature
onto the Frenet-Serret orthonormal basis $\vct{T}$, $\vct{N}$, $\vct{B}$.}}
\end{figure*}

\section{Characterization of a propagating wave packet}

{Visualization of the overall reconnection phenomenon, as it is displayed in
Fig.~\ref{fig:RepreReconn} suggests the creation of a localized phenomenon
along the vortices. To exhibit this phenomenon quantitatively, we
display in Fig.~\ref{fig:Soliton} maps of local curvature, i.e. the
numerical estimation, thanks to the vortex tracking algorithm, of
spatiotemporal maps of $\kappa(s,t)$ as a function of the arc-length $s$ and
time $t$. We indeed observe in both the local and non-local cases the
propagation of a wave packet of curvature, as it is predicted by the LIA, and
conveniently formalized in Ref.~\cite{Has72}. As far as the local case is concerned, we study moreover two particular values for the interaction parameter $g$ entering in the dynamics (Eq. \ref{eq:GPL}). As we have seen in Section \ref{sec:NumInv}, $g$ governs in the same time the speed of sound (i.e. $c_s=\sqrt{2g}$) and the healing length (i.e. $\xi=1/\sqrt{g}$) once is chosen $g=\mu$ as it is done in Eq. \ref{eq:GPL}. Thus, speed of sound is inversely proportional to the healing length. In our situation, the value $g=128$ (the respective curvature map and related quantities are displayed in Figs.~\ref{fig:Soliton}(a) to (d)) ensures a same speed of sound as in the non local version (Eq. \ref{eq:GPNL}), but with a vortex core diameter smaller than what is obtained in the model with rotons (see in particular Fig. \ref{fig:RadialDistrib}(b)). We also study the value $g=21.345$ (displayed in Figs.~\ref{fig:Soliton}(e) to (h)) for which vortex core size is similar to the nonlocal case (data not shown) but with a smaller speed of sound. Respective results concerning the non local case are displayed in Figs.~\ref{fig:Soliton}(i) to (l). As we will see in the following, the propagation of a wave packet is clear in the $g=21.345$ and in the non local cases, whereas it is less clear in the $g=128$ case. This might be due to the smallness of vortex cores that makes them stiff and somehow imposes a much higher speed of propagation of the wave packet, which turns out to be difficult to track.  Remark first that indeed, the
wave packet, and in particular the local maximum of curvature (i.e. bright colors),
seems to propagate approximately at a constant speed i.e. $c =
ds/dt$, for both the local and non-local cases, although in the non-local case,
secondary maxima appear during the propagation. }

To quantify more accurately the wave packet velocity and its shape in the vicinity
of the curvature maximum, we display in Figs.~\ref{fig:Soliton}~(b), (f) and (j) a
tentative rescaling of the observed curvature at various instants as 
\begin{equation}\label{eq:ABcSoliton}
\gamma(t)\kappa \left(\frac{s-c(t)t}{\sigma(t)},t\right),
\end{equation}
where $c(t)$ is the velocity of the maximum of curvature, $\sigma(t)$ a typical
length quantifying the increase in the width of the {wave packet}, and $\gamma(t)$
that allows to include a possible time-variation in the amplitude. We indeed
observe that this rescaling procedure makes the curvature profile similar to
what is observed initially. Under the LIA, it is shown in Ref.~\cite{Has72}
that such a {wave packet} is expected to behave as a solitary wave that propagates at
a constant velocity (given by the initial torsion) and does not change neither
its shape, nor its amplitude, i.e. $\sigma(t)=\gamma(t) = 1$, the actual time
independent shape being given by a hyperbolic secant function. {In the sequel, we will mostly focus on the $g=21.345$ local case, and on the non local case, since for the $g=128$ local case, the propagation of the wave packet appears to be different, and compares poorly with predictions of LIA.} In our present
numerical simulation, we see that in a good approximation, except at early
times ($t<0.1$), the velocity of the {wave packet}, $c(t)$, is nearly constant, 
of order $c\approx 1$ in the
local case (Fig.~\ref{fig:Soliton}(g)), and of order $c \approx 2$ (Fig.~\ref{fig:Soliton}(k)) in the non-local one, as it is suggested in
the LIA approach. In comparison, the celerity of sound in the superfluid is
$c_s=16$ in these units: the observed {wave packet} propagates in a much slower way than
acoustic waves, {although in the local stiff case (Fig.~\ref{fig:Soliton}(c)), the velocity of the wave packet may reach such values}. On the contrary, the {wave packet} undergoes both dispersion, i.e. 
its width increases as tracked by the increase of the rescaling coefficient
$\sigma(t)$, and its amplitude decreases, with accordingly a decrease in the
coefficient $\gamma(t)$. A more precise analysis shows indeed that
$\gamma(t)\sim 1/t$, a decrease that is not predicted by LIA.

{Focusing on the local $g=21.345$ and non local cases, in order to interpret these observed behaviors, we display in 
Figs.~\ref{fig:Soliton}~(h) 
and (l) the projections of the vortex velocity vector
$\partial \vct{X}/\partial t$ in the Frenet-Serret frame of reference
$(\vct{T}, \vct{N}, \vct{B})$ as a function of time. We indeed observe that the
projection along the binormal vector is constant during the {wave packet}
propagation, as it is assumed in the LIA. Let us precise here that if indeed
the projection of $\partial \vct{X}/\partial t$ on $\vct{B}$ appears to be time
independent, we can infer from the tracking of the {wave packet} itself 
(Figs.~\ref{fig:Soliton}~(h) and (l)) that the actual value of this projection cannot
be given by only the curvature since curvature itself is found dependent on
time. Interestingly, at early stage following reconnection, once again for
$t<0.1$, we see a non vanishing contribution along the normal vector
$\vct{N}$. As shown in Ref.~\cite{KleMaj91}, this part of the dynamics is
involved in a self-stretching phenomenon that would modify the length of
vortices, as it is indeed observed in Fig.~\ref{fig:LengthApp}~(a). As time
evolves, this projection gets smaller and smaller. Finally, we see that in the
local case, the projection along the tangential vector $\vct{T}$ is always
negligible in front the projections along the other direction of the frame,
whereas it cannot be neglected in the non-local case. This might be explained
while studying the interaction with additional local maxima that appear during
the {wave packet} propagation for the non-local case. Returning to the stiff $g=128$ local case (Fig.~\ref{fig:Soliton}~(d)), once again we see that the projection on the binormal vector remains also constant, and the projection on the normal vector remains negligible. On the contrary, the projection on the tangential vector may take large negative values, a phenomenon that remains to be understood. Let us mention again that since the wave packet travels very fast, it is difficult to track and the strong variations observed on the projection along the tangent vector may be due to the existence of several packets which interact in a complicated way.}

\section{Conclusion and final remarks}

We have studied numerically the reconnection phenomenon of two initially
orthogonal quantum vortices in a model of superfluids which includes the roton
minimum in the dispersion relation. As it was proposed in the 
literature~\cite{PomRic93,BerRob99,Ber99}, such a model describes the dynamics of a scalar
wave function that is governed by the Gross-Pitaevskii equation~\cite{PitStr03}
where is considered a non-local two-body interaction of characteristic extension
the $^4\mathrm{He}$ atomic size $a$. We start by calibrating the model to be as close as possible to the experimental dispersion relation of $^4\mathrm{He}$ provided in 
Ref.~\cite{DonBar98}, with the additional will  to prevent from the generation of
precursors of crystallization, as it was evidenced in Ref.~\cite{Ber99},
although of great importance in the context of supersolidity~\cite{PomRic94,JosPom07}. 
Once obtained such a model, we estimate its
stationary vortex solution, that is a time independent solution with an axis
symmetry, around which the phase turns of an angle equals to 2$\pi$. Then, in a
similar way as in Ref.~\cite{KopLev93}, we prepare a initial condition made up
of two of these vortices in a orthogonal situation. We indeed observe a
reconnection, and track and study the time evolution of the density, current
and pseudo-vorticity fields, we provide both a statistical analysis of the
fields and a local estimation of the geometry of these vortices, including a
precise characterization of a observed {wave packet} that shares some features
predicted by the LIA\@.

This numerical investigation shows that taking into account a more realistic
structure of vortices, as depicted in Ref.~\cite{VilCas12}, has little
influence on the global picture of reconnection given by the local version of
the GP equation and presented in Ref.~\cite{KopLev93}, although the creation
and propagation of a {wave packet} of curvature along the vortex cores appear more
complex when the non-local interaction is plugged in the dynamics.

The statistical analysis of the kinematic quantities involved in the dynamics,
in particular current $\vct{j}$ and pseudo-vorticity
$\vct{w}=\vct{\nabla}\wedge\vct{j}$, shows that there is no creation of scales
smaller that the injected atomic length size $a$. This makes a big difference
with what is obtained with the incompressible Euler or Navier-Stokes equations,
where a cascading phenomenon transfers energy towards the small scales, as
recently put in evidence while considering two colliding vortex rings in a
experimental (classical) flow~\cite{KeoMon18}. We can thus infer that the
hydrodynamics implied by the local and non-local versions of the GP equations,
because of its implied high level of compressibility in the vicinity of the
vortices, and the unclear action of the additional quantum pressure term, is
indeed very different from the one of incompressible viscous Newtonian fluids. 

It would of tremendous importance to develop an interaction term in the GP
evolution of the wave function able to include a more realistic prediction of
the dispersion relation, without exhibiting crystallization phenomena that are
not expected in the superfluid phase of $^4\mathrm{He}$. We keep this perspective for
future investigations.

\begin{acknowledgments}
We thank B. Castaing for numerous fruitful discussions and with who we started
this project, and the PSMN for computer time. Authors are partially
founded by ANR Grants No. LIOUVILLE ANR-15-CE40-0013 and No. ECOUTURB ANR-16-CE30-0016. 
\end{acknowledgments}


%

\end{document}